\documentclass[useAMS,usenatbib]{mn2e}
\usepackage{graphicx}
\usepackage{epstopdf}
\usepackage{fixltx2e}
\usepackage{color}

\title[The $\lambda$6614 diffuse interstellar absorption band: evidence for internal excitation of the carrier]{The $\lambda$6614 diffuse interstellar absorption band: evidence for internal excitation of the carrier}
\author[Charlotte C. M. Marshall, Jacek Kre{\l}owski and Peter J. Sarre]{Charlotte C. M. Marshall$^{1}$\thanks{E-mail:
pcxccm@nottingham.ac.uk}, Jacek Kre{\l}owski$^{2}$\ and Peter J. Sarre$^{1}$\\
$^{1}$School of Chemistry, The University of Nottingham, University Park, Nottingham NG7 2RD\\
$^{2}$Centre for Astronomy, Nicholas Copernicus University, Gagarina 11, Pl-87-100 Toru$\acute{n}$, Poland}

\begin{document}

\date{Accepted ???. Received ???; in original form ???}

\pagerange{\pageref{firstpage}--\pageref{lastpage}} \pubyear{2015}
\maketitle

\label{firstpage}

\begin{abstract}

An analysis of absorption profiles of the $\lambda$6614 diffuse interstellar band recorded along the lines-of-sight towards HD~179406 (20~Aql) and HD~147889 is described. The difference in band shape is attributed to the degree of internal excitation of the carrier, which is principally due to vibrational hot bands although an electronic component may also be present.  The results are discussed with respect to other models of diffuse band spectral line shape.


\end{abstract}



\section{Introduction}
\label{sec:intro}
Identification of the carriers of the enigmatic Diffuse Interstellar Bands (DIBs) is often referred to as the biggest current challenge in astronomical spectroscopy. The observational data comprise over 500 interstellar absorption features, predominantly at visible wavelengths, which are observed towards bright O and B stars seen though significant foreground material.  Although this field of research was established nearly one hundred years ago, efforts to identify the carriers over the years have not met with success. However, in a significant development, based on the results of gas-phase laboratory experiments, \cite{Camp2015} conclude that they have positively identified two near-infrared diffuse interstellar absorption bands as due to transitions of C$_{60}^+$. The overall subject area has been reviewed by \cite{Herbig1975} and \cite{Sarre2006}.

Some of the diffuse bands have been known to exhibit distinct shapes for at least forty years \citep{Herbig1975,West1988a,West1988b}. Diffuse band substructure in the form of a triplet was first observed in the $\lambda$6614 band by \cite{Hibbins1994} towards $\mu$\,Sgr, $\sigma$\,Sco, $\beta^1$\,Sco and $\zeta$\,Oph. Using the Ultra-High-Resolution-Facility (\textit{RP} = 600,000) on the Anglo-Australian Telescope, the diffuse bands {$\lambda$}{$\lambda$}6614 and 5797 were found to have distinct components within the profiles, which provides support for a molecular rather than a condensed-phase origin \citep{Sarre1995, Kerr1996, Kerr1998}. Rotational contour modelling of the $\lambda$6614 triplet profile towards $\mu$\,Sgr using R, Q and P-type rotational branches yielded planar oblate symmetric top molecules of polycyclic aromatic hydrocarbon (PAH) or cyclic form as potential carriers \citep{Kerr1996}.  It was noted by \cite{Kerr1996} that an additional absorption found in spectra of $\lambda$6614 towards $\mu$~Sgr was not accounted for in `R, Q, P' triplet interpretation or by the contour modelling, but that it might be explained in terms of `vibrational hot bands, isotopic modifications or splitting of the level structure by magnetic (e.g. spin-orbit) or Jahn-Teller interactions'.  An additional feature labelled `c' that appears on the redward side in spectra of $\lambda$5797 \citep{Kerr1998} probably has a similar origin.

Triplet and doublet structure for $\lambda$6614 and $\lambda$5797, respectively, was reported by \cite{Ehren1995} at \emph{RP}~=~70,000 as well as a doublet for $\lambda$6379; the separation between `R' and `P' peaks enabled estimates of rotational constants of the carriers to be obtained which are consistent with carriers being PAH molecules with more than 40 carbon atoms, 12-18 carbon chains or compounds of C${_{60}}$ \citep{Ehren1996}.  This was further developed by \cite{Cami2004} who noted variation in the relative positions of the peaks within $\lambda$6614 according to sight-line, which could be interpreted in terms of the rotational excitation temperature of a molecular carrier.  An alternative interpretation for the multiplet structure has been offered by Webster (\citeyear{Webster1996}) in which the substructure peaks are attributed to \textsuperscript{13}C isotopic modifications of the carrier.

Newly discovered extended red tails for some of the diffuse bands recorded through an infrared-illuminated region towards Herschel~36 have also been interpreted in terms of molecular rotational contours \citep{Dahlstrom2013, Oka2013}. In their analysis the extended contours of $\lambda$$\lambda$5780, 5797 and 6614 are considered to be due to polar molecules with three to six heavy carbon-like atoms. In contrast, $\lambda$$\lambda$5850, 6196 and 6379, which do not have pronounced tails, were suggested to be likely due to non-polar molecules or large polar molecules with a low relative change in rotational constant on electronic excitation.  For both contour types only a single vibronic band was included in the calculations.  While attributing the contour shapes to single rotational contours of relatively small molecules, the authors did not rule out other interpretations including the role of low-frequency vibrations.

In a recent study \cite{Bernstein2015} concluded that the overall profiles of $\lambda$6614 detected in diffuse and translucent clouds result from accidentally overlapping diffuse bands centred at 6613.6~\AA\ (the strongest component of the triplet) and at 6614.2~\AA\ (a single highly asymmetric red degraded band), where these independent bands are due to different carriers - based on a suggestion made earlier by \cite{Galazutdinov2002}.  This interpretation differs markedly from the internal excitation `hot band' origin as put forward by \cite{Kerr1996} which involves a single carrier and is further considered and modelled here.

Particularly in the light of high-quality HARPS data we have re-examined the hot-band suggestion, including line-of-sight profile comparison, co-addition of principal and hot band profiles, and molecular rotational contour calculations. We describe a model in which the absorption on the redward side of $\lambda$6614 for \emph{e.g.} HD~147889 -- which is enhanced in comparison with HD~179406 -- is attributed to `hot band' contributions, as would be expected to arise due to a higher internal `temperature' in a molecule with low-lying vibronic states.   For simplicity we choose to use the term `hot band' irrespective of the precise vibrational, electronic etc. origin of the excited level(s) which is unknown.

\section{Diffuse band profiles and molecular absorption features}
\label{sec:molabs}
In diffuse clouds the rotational temperature of optically detected polar interstellar molecules such as CN is generally close to that of the cosmic microwave background radiation temperature of 2.72548~$\pm$0.00057~K \citep{Fixsen2009}, though is generally found to be slightly higher, probably due to the effect of electron collisions \citep{Black1991,Ritchey2011,Krelowski2012,Leach2012,Harrison2013}.  A notable exception is the line-of-sight towards Herschel~36 with rotational temperatures of 6.7~K and 14.6~K determined for CH and CH$^+$, respectively.  The highly unusual display of diffuse band shapes most probably arises from a higher-than-usual degree of diffuse band carrier internal excitation.
Electronic absorption spectra of non-polar molecules in diffuse clouds including H$_2$, C$_2$ and C$_3$ carry information on the ground-state rotational level populations which are determined by the balance between collisional and radiative excitation processes; effective rotational `temperatures' of up to \emph{c.} 200~K have been found.  However, the physical interpretation of the rotational level column densities is not straightforward as discussed by \cite{Casu2012}, \cite{Hupe2012}, \cite{Welty2013} and in citations of earlier work.

Observations of variation in diffuse band shapes include both blue and red shifts and band broadening where these have been shown to be unrelated to Doppler effects from thermal or turbulent motion or due to multiple line-of-sight clouds.  Early examples include $\lambda$5780 \citep{Porceddu1992} for which unusually large widths were found towards three stars in Orion, and small red shifts for $\lambda$$\lambda$6614, 5780, 5797, 6284 and 5705 towards HD~37022 and HD~37061 which are also in Orion \citep{Krelowski1999}. Comparative blue shifts have also been reported for HD~37048 and stars in the Sco~OB1 association \citep{Gal2006,Gal2008,Gal2015}. High-resolution HARPS spectra have revealed marked differences in diffuse band profiles, for example towards HD~179406, HD~163800, HD~147165 and HD~147889 \citep{Galazutdinov2008highres}.  A good overall summary of studies of diffuse band profile variation is presented in section 4.2 of \cite{Dahlstrom2013}.

\section{Observational data}
\label{sec:sightlines}

The star HD~179406 (20 Aql) lies behind a $\zeta$-type interstellar cloud (W(5797)/W(5780)= 0.48, E$_{B-V}$=0.31), as does HD~147889 (W(5797)/W(5780)= 0.42, E$_{B-V}$= $\sim$1). Of particular importance, as inferred from the velocity profiles of the K\,\textsc{i} and CH absorption lines, there is a single dominant velocity component and hence very likely a single interstellar cloud in each case \citep{Galazutdinov2008highres}. However, this does preclude the presence of regions with differing velocity dispersions or conditions within the cloud as the spectrum recorded is integrated along the line-of-sight. The profile of $\lambda$6614 is known to differ significantly with line-of-sight - see fig~3. of \cite{Galazutdinov2008highres}, both in band width and in the relative strengths of the triplet components. The observational data considered here were obtained by \cite{Galazutdinov2008highres} using the High Accuracy Radial Velocity Planet Searcher (HARPS) \'{e}chelle spectrograph, with sightlines selected where possible based on their having a single CH\,4300~\AA\ line with low velocity dispersion \citep{Galazutdinov2008highres}. The laboratory rest frequency for the carrier of $\lambda$6614 is obviously not known and it is assumed that it shares the same radial velocity shift as for CH as diffuse band strengths are generally found to correlate well with the column density for this molecule, when observed.

\section{Profiles of $\lambda$6614 towards HD~179406 and HD~147889}
\label{sec:profiles}

\subsection{Profile comparison}

The high resolving power (\emph{RP}~$\sim$~115,000) of the HARPS instrument means that the substructure and shapes of the individual components can be seen in some detail as shown in the upper panel of fig~1. In this figure the two bands are constrained to have a common normalised integrated intensity (equivalent width).\footnote{In the molecular interpretation a diffuse band comprises many unresolved rotational lines each with their own quantum mechanical line strength, and as discussed here, can also include hot bands -- each with their own Franck-Condon factor. It follows that when comparing spectra arising from carriers with different levels of internal excitation the diffuse band equivalent widths are not necessarily exactly proportional to the carrier column densities. Comparing normalised profiles in this way is of value for discussion purposes; it is no less rigorous than normalising to the strongest peak in the profiles.}  This reveals that the band towards HD~147889 is less deep but is broader than that recorded towards HD~179406.

The basis for our interpretation is that the main triplet structure arises from R, Q and P rotational branches of the origin band of a molecular electronic transition \citep{Sarre1995}.  Considering the overall profiles and starting from an onset at \emph{c.} 6612.9~\AA, there is a sharp increase in absorption which suggests the presence of a rotational (R) branch head. The second and always strongest feature (Q) has a less steep long-wavelength side towards HD~147889. For HD~179406 a third possible (P) shoulder near $\sim$6613.8~\AA\ is quite weak; this is a common characteristic of partially resolved red-degraded molecular rotational profiles. However, for HD~147889 this third feature is quite narrow and strong, particularly when comparing its intensity relative to the R and Q features towards HD~179406; the relative strengths of the first (R) and third (`P') peaks are in fact reversed between these two lines-of-sight. Finally, the redward extension of $\lambda$6614 for HD~147889 is significantly stronger than for HD~179406 and a fourth narrow absorption feature at 6614.2\,\AA\ is far more prominent towards HD~147889.  Other examples of $\lambda$6614 profiles are given in \cite{Galazutdinov2002}.

 The difference between the spectra towards HD~179406 and HD~147889 is presented in the lower panel of figure~1. To a first approximation this trace reflects the change in internal population distribution of the $\lambda$6614 carrier between the two clouds; the intensity of the R feature of the origin band is reduced towards HD~147889 and the Q maximum is also reduced but less so, the third `P' peak is more prominent towards HD~147889, and additional absorption in the 6613.7 - 6615.5\,\AA\ region is evident. These observations are consistent with partial redistribution of the ground-state ro-vibrational population for HD~179406 to one or more excited levels for the carrier towards HD~147889, where these levels must lie close in energy to the ground vibrational state in order for this to be of relevance under interstellar conditions.

\begin{figure}
  \begin{center}
     \includegraphics[width=\hsize]{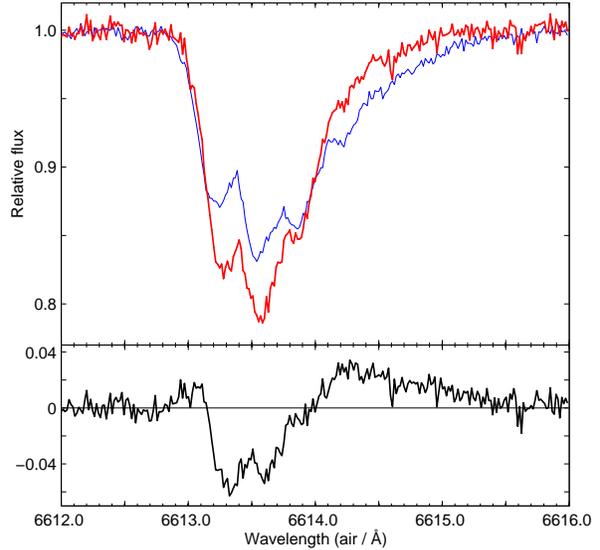}
  \end{center}
  \caption{Upper panel - profiles of $\lambda$6614 towards HD~179406 (deeper, thick red trace) and HD~147889 (broader, thin blue trace). The spectra are normalised to a common integrated intensity and the ordinate values are those for HD~179406. Lower panel - difference between HD~179406 and HD~147889.}
  \label{fig:normal}
\end{figure}

\subsection{Co-addition of profiles}
\label{sec:co-addition}

In attributing the changes in profile between HD~179406 and HD~147889 to differences in the quantum level populations of the carrier, we suggest that the single largest effect is the role of hot bands, probably vibrational in origin, with an additional contribution from a small change in rotational temperature. This is illustrated in fig~2. where the top two panels show the spectra towards (a) HD~179406 and (b) HD~147889 for reference.  In fig~2(c) addition of a wavelength-shifted weaker copy of the profile towards HD~179406 to the standard profile of $\lambda$6614 for HD~179406, yields a profile which is a good approximation to that observed towards HD~147889.  This treats the shift and the relative weight of the origin and hot band as free parameters. There are small discrepancies in the region of the R head and in the longer wavelength part of the red tail.  The first discrepancy may be explained as due to a higher rotational temperature for HD~147889 which results in an R-branch which extends to a slightly shorter wavelength than for HD~179406; the second difference is largely accounted for through addition of a second weaker hot band, the result being shown in the lowest panel (fig. 2 (d)). The hot band shifts reflect differences in the spacings of the low-frequency intervals in the ground and excited electronic electronic states. Inclusion of further weak hot band contributions could achieve a yet better fit to the tail in a technical sense, but there is no certainty that only one vibrational mode contributes, and the co-addition procedure implicitly assumes that the rotational constants and rotational populations -- and therefore rotational profiles, do not change either between vibronic level or from those for HD~179406.  In Section~\ref{sec:modelling} we show that there is likely a small contribution to the profile from a hot band even for HD~179406.  Notwithstanding these limitations, we find the fact that simple co-addition of shifted copies of the HD~179406 profile accounts for most of the profile towards HD~147889 as in fig. 2 to be supportive of the proposed hot band model.

An excellent laboratory example of the behaviour of low-frequency hot bands in electronic spectra is seen in a jet-cold spectrum of the cyclic C$_{18}$ molecule, where evidence for additional vibrational bands both overlapping and to the redward side of the origin band was found by varying the sample temperature \citep{Maier2006}.  Analogous astrophysical examples include radiation-pumped population transfer to excited vibrational and rotational levels of H$_2$, and observed non-thermal population distributions for high rotational levels of C$_2$ and C$_3$ which are found to vary with line-of-sight.  However, not all reddened lines-of-sight with diffuse band features including $\lambda$6614 have detectable C$_2$ or C$_3$.

\begin{figure}
  \begin{center}
     \includegraphics[width=\hsize]{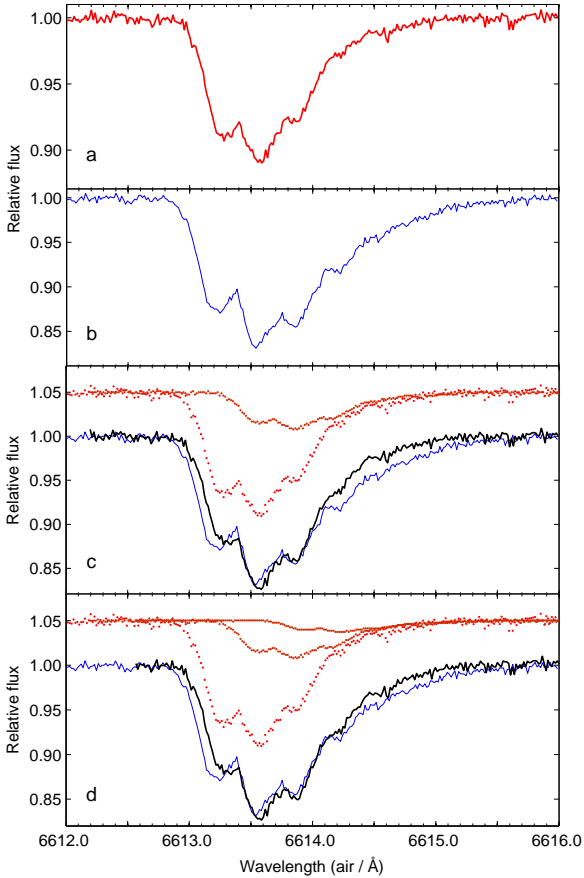}
  \end{center}
  \caption{Upper two panels -- $\lambda$6614 towards (a) HD~179406 (red) and (b) HD~147889 (blue).  Third panel (c) -- profile recorded towards HD~179406 (weight 0.77), a weaker red-shifted copy of HD~179406 (shift -0.66\,cm$^{-1}$) with weight 0.23, and the result of co-addition (thick black trace) compared with the profile towards HD~147889 (thin blue trace). Bottom panel (d) -- as for (c) with addition of a second shifted copy of HD~179406 (shift -1.51\,cm$^{-1}$) with  weights of 0.72, 0.22 and 0.06.}
  \label{fig:co-addition}
\end{figure}

\section{Contour modelling including hot bands}
\label{sec:modelling}

\subsection{Background}
Rotational contour modelling of diffuse band profiles was first conducted by \cite{Danks1976} and has become a popular way in which to seek to narrow the range of possible diffuse band carriers.  It was first applied to $\lambda$6614 for $\mu$\,Sgr \citep{Kerr1996} and further applications to this diffuse band involve cumulenes \citep{Schulz2000}, contaminated H$_2$ clusters \citep{Bernst2013}, small linear molecules \citep{Oka2013} and accidentally overlapping diffuse bands including a prolate top \citep{Bernstein2015}.

\subsection{Contour modelling}
\label{sec:modelling}

While the co-addition approach of section 4.2 is a useful first step in exploring the impact of vibrational hot bands on diffuse band profiles, it is preferable to model the profiles of $\lambda$6614 towards HD~179406 and HD~147889 with molecular rotational contours.  It was already demonstrated that a planar oblate symmetric molecule could provide a good match to the profile towards $\mu$\,Sgr \citep{Kerr1996} and we make use of this here, but now including vibrational hot band structure.  This was undertaken initially for HD~179406 as it appears to be the simplest and most likely the coolest spectrum.  The model adopted is that reported by \cite{Kerr1996} for a vibronically allowed $\tilde{A}$$^{1}$B$_{2u}$~--~$\tilde{X}$$^{1}$A$_{1g}$ transition of a planar oblate symmetric top about the size of the coronene molecule. The same molecular parameters -- B$''$~=~0.00947\,cm$^{-1}$, $\Delta$B~=~-0.42\% and Coriolis parameter $\zeta$~=~-0.43 were employed and constrained to these values, while the rotational temperature and line width were allowed to vary in a PGOPHER fit \citep{Western2014}.  The fitting result for a single vibronic band for HD~179406 is shown in fig.~3(a) where the parameters deduced are: Gaussian line width FWHM~=~0.43(1)\,cm$^{-1}$, T~(band origin, air)~=~15,120.45(1)\,cm$^{-1}$ and T$_{rot}$~=~18.5(3)\,K.
The FWHM line width of 0.43\,(1)~cm$^{-1}$ corresponds to a velocity dispersion of 8.5\,~kms$^{-1}$ which is slightly higher than the value of 5.65\,kms$^{-1}$ reported for CH by \cite{Kazmierczak2009}. Inclusion of a very weak hot band results in a slight improvement to the fit (fig 3(b)) and suggests that there is a small contribution from a hot band even for HD~179406.

The molecular constants employed in the fitting of the HD~179406 data were constrained to these values and then used in fitting the spectrum recorded towards HD~147889; a very small shift in the origin of +0.08\,cm$^{-1}$ (1.6~kms$^{-1}$) was needed but falls well within the FWHM velocity profile of the (reference) CH spectrum with FWHM of 5.65~kms$^{-1}$ for HD~147889 \citep{Kazmierczak2009}.  Unsurprisingly, given the co-addition results for HD~147889 shown in fig. 2, hot band contributions are essential and the result of PGOPHER fitting is shown in fig. 4(a) (upper panel); the parameters floated were the rotational temperature, the line width and the weighting of two hot bands.  The rotational temperature is found to be slightly higher than for HD~179406 with a value of 22.0(2)~K. The longer wavelength region is generally well accounted for and while the overall resultant fit is not quite as good on the long-wavelength side as that attained by simple co-addition of copies of HD~179406 (fig. 2) described in Section~\ref{sec:co-addition}, this probably reflects deficiencies in the molecular rotational contour model.  These range from the choice of molecular geometry and electronic state symmetries to the non-inclusion of centrifugal distortion effects and use of a single rotational temperature across all J and K values. However, the major change in the relative strengths of the three main `triplet R, Q and P' absorption peaks between HD~179406 and HD~147889 is well accounted for.  The relative weights of the three vibrational bands (origin, first and second hot band) for HD~147889 were found to be 0.75\,:\,0.20\,:\,0.05 which is in accord with an exponential Boltzmann fall-off law for thermally populated equally spaced non-degenerate vibrational levels.  The line width determined for HD~147889 is FWHM~=~0.43(1)\,cm$^{-1}$.

While an exponential fall-off in the relative weight of the hot bands is an obvious starting point, the situation is complicated by the fact that (i) the level populations are not necessarily thermal and are probably determined by a blend of collisional and radiative excitation processes, (ii) the excited levels may in fact be electronic (\emph{e.g.} Jahn-Teller) rather than vibrational components -- or some combination of the two, (iii) the Franck-Condon factors of different hot bands are not expected to be equal, and (iv) the electronic transition moment may vary between hot band components.  Allowing inclusion of a third hot band and a weighting of 0.66:\,0.14:\,0.10:\,0.10\, yields an improved profile match as shown in fig 4.(b).

A non-thermal population distribution among vibrational (and rotational) levels might be expected for a non-polar planar oblate top molecule given data from rotationally resolved electronic absorption spectra of non-polar molecules such as C$_2$ and C$_3$ along the same lines-of-sight. Rotational populations for C$_2$ (X$^1\Sigma_g^+$) are available for both HD~179406 and HD~147889 \citep{Kazmierczak2010} and are presented in fig~5. In neither case is the plot linear which would hold for thermal population of rotational levels, and high rotational levels in particular are significantly populated for HD~147889.  Comparisons of diffuse band profile characteristics with these populations for C$_2$ has to be treated with some caution as some lines-of-sight such as the $\sigma$ line-of-sight HD~147165 ($\sigma$~Sco) have readily detectable $\lambda$6614 absorption but C$_2$ falls below the detection limit.  Nevertheless, to the extent that the $\lambda$6614 carrier and the C$_2$ molecule share some common spatial regions and conditions, it would appear that the `hotter' profile of $\lambda$6614 towards HD~147889, as inferred in this paper, correlates with population of the higher J levels in C$_2$.  Given that the higher-J tail for C$_2$ is normally attributed to radiative pumping, this suggests that the population of excited vibronic levels of the ground state of the $\lambda$6614 carrier towards HD~147889 originates in the same way.

\begin{figure}
  \begin{center}
   \includegraphics[width=\hsize]{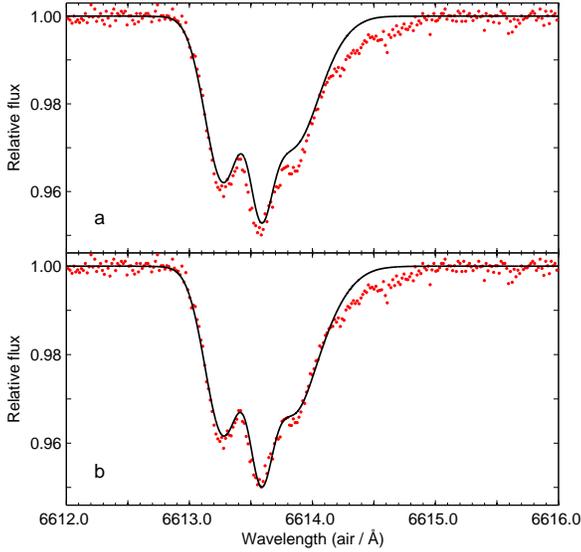}
  \end{center}
  \caption{Upper panel (a) -- $\lambda$6614 towards HD~179406 compared with result of PGOPHER modelling with no hot band included. Lower panel (b)  - as for (a) but with one hot band included with a redward shift of -0.66~cm$^{-1}$ and relative weight of 0.07.}
  \label{fig:179406model}
\end{figure}

\begin{figure}
  \begin{center}
    \includegraphics[width=\hsize]{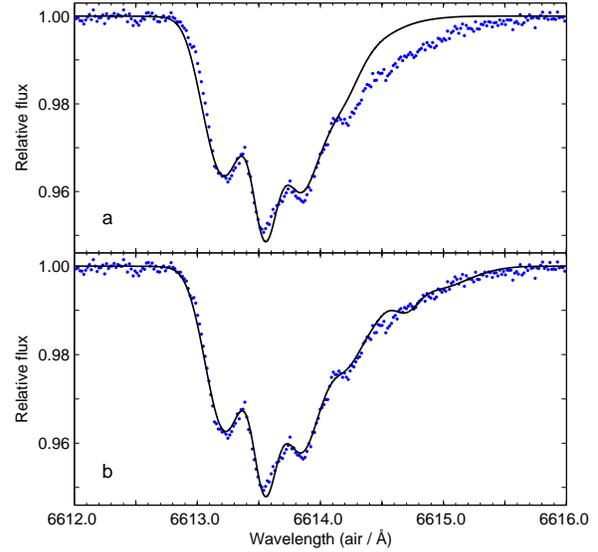}
  \end{center}
  \caption{Upper panel (a) -- $\lambda$6614 towards HD~147889 compared with result of PGOPHER modelling with two hot bands included (redward shifts of  -0.66~cm$^{-1}$ and -1.54~cm$^{-1}$) and a Boltzmann fall-off in hot band weightings of 0.75\,:\,0.20\,:\,0.05.  Lower panel (b) -- result of fitting with floated weights for three hot bands with shifts of -0.66~cm$^{-1}$, -1.54~cm$^{-1}$ and -2.64~cm$^{-1}$ with weights of 0.66:\,0.14:\,0.10:\,0.10\,(see text).}
  \label{fig:147889model}
\end{figure}

\begin{figure}
  \begin{center}
     \includegraphics[width=\hsize]{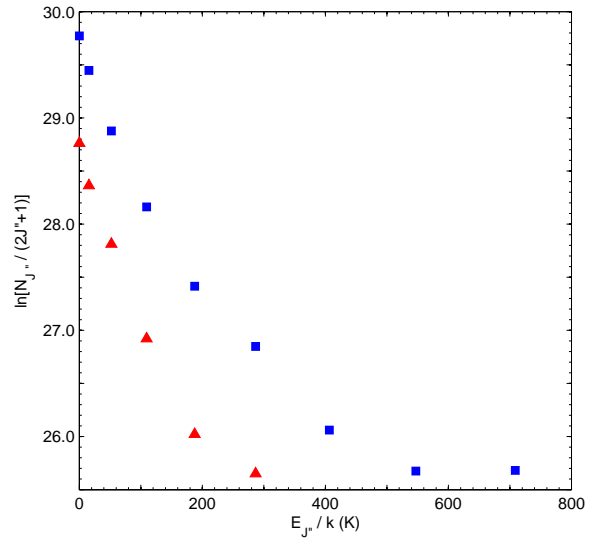}
  \end{center}
  \caption{Rotational level populations of the C$_2$ molecule towards HD~179406 (triangles, red) and HD~147889 (squares, blue) \citep{Kazmierczak2010} illustrating non-linear non-thermal distributions with an extended high-J tail for HD~147889.}
  \label{fig:C_2populations}
\end{figure}

\section{Summary and Conclusions}
\label{sec:concl}

We have conducted an investigation into the origin of the $\lambda$6614 fine structure, its overall band shape and its change between the lines-of-sight, HD~179406 and HD~147889. The interpretation offered is founded in terms of the population distribution among the rotational, vibrational and potentially low-lying electronic states of a medium-sized (N$_C$~$\sim$~20) planar PAH-type molecule.  The $\lambda$6614 carrier in the line-of-sight towards HD~147889 is found to have significantly higher internal excitation which accounts for its greater width and additional structure.  We consider that this may be a general feature of diffuse band profiles, but one that clearly depends on the presence, or otherwise, of suitable low-lying vibrational (or vibronic) states.  Moreover, the extent to which hot bands are evident in diffuse band spectra will depend on whether these transitions are significantly wavelength-shifted or simply overlap the origin band. However, in cases where no low-lying (populated) vibronic levels exist, slightly altered absorption profiles may still occur due to changes in rotational contours with temperature.

This interpretation differs from others recently put forward which consider the $\lambda$6614 profile in terms of a single rotational contour \citep{Oka2013} or invoke accidentally overlapping diffuse bands \citep{Bernstein2015}.  In the study of Herschel~36 by \cite{Oka2013} an extraordinary variety of diffuse band profiles were found which were grouped broadly in terms of the polarity of the band carrier, those with pronounced red tails including $\lambda$6614, $\lambda$5797 and $\lambda$5780 being attributed to polar molecules, with high rotational levels being populated due to irradiation by the nearby IR source Her~36~SE.  In the interpretation presented in this paper the extended red tails would arise principally from vibrational hot bands that are red-shifted with respect to the origin band, and bands without extended red tails would be examples where the carrier does not possess low-lying vibronic states and/or any hot bands present are not significantly wavelength-shifted from the origin band.

One aspect arising from these considerations is that the measured equivalent widths of a diffuse band such as $\lambda$6614 along different lines-of-sight do not necessarily give the exact column densities of the carrier.  This arises because transfer of population between levels which have different intrinsic transition strengths (\emph{e.g.} Franck-Condon factors for excitation from the lowest compared with excited vibrational levels) can result in weaker (or stronger) absorption for the same carrier column density.  In this context it is of interest to investigate whether this might play a role in the close, but not perfect, correlation of $\lambda$6614 and $\lambda$6196 \citep{Moutou1999,Galazutdinov2002,McCall2010}, the results of which will be reported elsewhere.

\section*{Acknowledgments}
We thank Michael Hollas and Steven Fossey for helpful initial discussions and Eloise Braun for assistance in reviewing the literature on laboratory electronic spectra of PAHs.  CCMM thanks EPSRC and The University of Nottingham for financial support.  PJS thanks the Leverhulme Trust for award of a Research Fellowship and Leiden Observatory for hospitality that allowed completion of this work. JK acknowledges the financial support of the Polish National Center for Science during the period 2012 - 2015 (grant UMO-2011/01/BST2/05399).

\bibliographystyle{mn2e}
\bibliography{bib}

\bsp

\label{lastpage}

\end{document}